\documentclass[preprintnumbers, floatfix, twocolumn,
preprintnumbers, PRD, superscriptaddress,nofootinbib]{revtex4}
\usepackage[utf8]{inputenc}
\usepackage{eurosym}
\usepackage{amsfonts}
\usepackage{amsmath}
\usepackage{amssymb,epsf}
\usepackage{latexsym}
\usepackage{graphicx,epsfig}
\usepackage{amssymb}
\usepackage{subfigure}
\usepackage{epstopdf}
\usepackage[colorlinks=true,citecolor=blue,linkcolor=blue,urlcolor=black]{hyperref}
\usepackage{color}

\usepackage{float}

\graphicspath{{Images/}}

\renewcommand{\&}{\textup{\symbol{`\&}}}

\begin{document}
\title{Lifshitz scaling effects on the holographic paramagnetic-ferromagnetic phase transition}

\author{B. Binaei Ghotbabadi}
\affiliation{Physics Department and Biruni Observatory, Shiraz
University, Shiraz 71454, Iran}

\author{A. Sheykhi}
\email{asheykhi@shirazu.ac.ir}

\affiliation{Physics Department and Biruni Observatory, Shiraz
University, Shiraz 71454, Iran}

\affiliation{Research Institute for Astronomy and Astrophysics of
Maragha (RIAAM), Maragha, P. O. Box: 55134-441, Iran}

\author{G. H. Bordbar} \email{ghbordbar@shirazu.ac.ir}
\affiliation{Physics Department and Biruni Observatory, Shiraz
University, Shiraz 71454, Iran}

\begin{abstract}
We disclose the effects of  Lifshitz dynamical exponent $z$ on the
properties of holographic paramagnetic-ferromagnetic phase
transition in the background of Lifshitz spacetime. To preserve
the conformal invariance in higher dimensions, we consider the
Power-Maxwell (PM) electrodynamics as our gauge field. We
introduce a massive $2$-form coupled to the PM field and perform
the numerical shooting method in the probe limit by assuming the
PM and the $2$-form fields do not back-react on the background
geometry. The obtained results indicate that the critical
temperature decreases with increasing the strength of the power
parameter $q$ and dynamical exponent $z$. Besides, the formation
of the magnetic moment in the black hole background is harder in
the absence of an external magnetic field. At low temperatures,
and in the absence of an external magnetic field, our result show
the spontaneous magnetization and the ferromagnetic phase
transition. We find that the critical exponent takes the universal
value $\beta= 1/2$ regardless of the parameters $q, z, d$, which
is in agreement with the mean field theory. In the presence of an
external magnetic field, the magnetic susceptibility satisfies the
Curie-Weiss law.
\end{abstract}


\maketitle

\section{Introduction}
The Bardeen, Cooper and Schrieffer (BCS) theory of
superconductivity is a well known microscopic theory for studying
the weakly coupled low temperature superconductors
\cite{Bardeen1,Cooper2}. However, this microscopic theory suffers
to explain the pairing mechanism of the materials with strongly
coupled interaction at high temperature. Therefore, it is
important to find an alternative approach for describing the high
temperature superconductors. The correspondence between a strongly
coupled conformal field theory (CFT) in $d$-dimensions and a
weakly coupled gravity theory in ($d+1$)-dimensional anti-de
Sitter (AdS) spacetime provides a powerful tool to shed the light
on the mechanism of high temperature superconductors
\cite{Maldacena1,Gubser2,Witten3}. Investigation of the electronic
properties of materials and magnetism by employing AdS/CFT duality
is one of the applications of this approach
\cite{montull,Donos,albash,m.pujo,iqbal,hartnoll,Herzog5,McGreevy6,Herzog7,Gubser8}.
The first holographic superconductor considers a four-dimensional
Schwarzschild-$AdS$ black hole coupled to a Maxwell and a scalar
fields in the simple model \cite{Building Hartnoll}. A variety of
the holographic dual models have been explored in the literatures
(see e.g. \cite{Lai,Rogatko,Kuang,Mansoori,Ling,Introduction
Cai,Lifshitz Wu,Bahareh, Mahya1,Mahya2,Mahya3,Mahya4} and
reference therein).

The holographic paramagnetic-ferromagnetic phase transition in a
dyonic Reissner-Nordstrom-AdS black brane is another example which
gives a starting point for exploration of complicated magnetic
phenomena and quantum phase transition \cite{dyonic}. This model
was extended by introducing two antisymmetric tensor fields which
correspond with two magnetic sublattices in the materials
\cite{p.Acai6}. In the framework of usual Maxwell electrodynamics,
the physical properties of holographic
paramagnetism-ferromagnetism phase transition have been
investigated by some authors
\cite{p.Acai6,Coexistence.Cai,Yokoi,Cai3,Cai4,Insulator.Cai,Understanding.Cai}.
Although in AdS/CFT the action should be considered as an
effective approach of string theory, in general, for
applications of gauge/gravity duality such as the holographic
superconductor model, the gravitational model could not be studied
as well as those which satisfy the behavior of boundary theory
and the condition of string theory. The employing different
electromagnetic actions can modify the dynamics of dual theory.
Since in this viewpoint, the nonlinear electrodynamic theories
correspond to the higher derivative corrections to the Abelian
gauge fields; therefore it can be useful for these kinds of
investigations. These nonlinear electrodynamics carry more
information than that of the Maxwell field, and they have been
interesting subjects of research in the recent years. As an
example, the effect of BI-like electrodynamics parameter on the
paramagnetism-ferromagnetism phase transition has studied in Refs.
\cite{Zhang2,Wu1}. Considering Born-Infeld-likes electrodynamics,
it has been observed that the higher nonlinear corrections make
the formation of magnetic moment being harder, decreasing the
critical temperature and changes the condensation gap in the
absence of external magnetic field. The properties of holographic
superconductor with conformally invariant PM electrodynamics have
been studied in Refs. \cite{PM1,PM2,Shey1,Shey2,Shey3,Shey4,Doa}.
Now it is interesting for us to investigate how the power-law
Maxwell field as another type of nonlinear electrodynamics,
affects the paramagnetism-ferromagnetism phase transition. Here,
One of the reasons is that the Power-Maxwell field takes the
special asymptotical solution near boundary which is different
from all known cases.

On the other hand, in the framework of condensed matter physics, a
dynamical scaling appears near the critical point with the scale
transformation turns to be \cite{Lif}
\begin{equation}\label{eq00}
t \rightarrow \lambda^{z} t, \ \    x_{i} \rightarrow \lambda
x_{i}, \  \  z\neq0,
\end{equation}
where $z=1$ corresponds to usual AdS spacetime. It was
pointed out that at the quantum critical point there is a Lifshitz
scaling similar to Eq. (\ref{eq00}) \cite{carlos}.  A lot of
attempts have been carried out in Lifshitz scaling by using the
holographic approach (see e.g.
\cite{Bu,LU,Natsuume,Sherkatghanad,Zhao14, Mahya5}). However, the
most of the previous works regarding the effects of Lifshitz
scaling on holographic superconductors were done by considering a
SU(2) Yang-Mills gauge field in the bulk in the presence of
Maxwell electrodynamics \cite{Bu,LU}. The holographic
paramagnetism-ferromagnetism phase transition in the four and
five-dimensional Lifshitz spacetime has been explored in
\cite{Lifshitz5} by introducing a massive $2$-form field coupled
to the Maxwell field. It has been confirmed that the Lifshitz
dynamical exponent $z$ contributes evidently to magnetic moment.
Holographic model for ferromagnetic phase transition in the
Lifshitz black hole in the presence of Born-Infeld-like nonlinear
electrodynamics has been investigated in Ref. \cite{Zhang2plb}.
It was observed that, in case of larger dynamical exponent $z$,
the exponential form of nonlinear electrodynamics correction leads
to the smaller value for critical temperature and the magnetic moment comparing with the logarithmic and Born-Infeld
types of nonlinear electrodynamics \cite{Zhang2plb}.

In our previous work \cite{binaei}, we explored the effects of PM
nonlinear electrodynamics on the properties of holographic
paramagnetic-ferromagnetic phase transition in the background of
Schwarzchild-AdS black hole. We wonder how other backgrounds
affect the paramagnetism-ferromagnetism phase transition,
especially the Lifshitz spacetime as an example of the
non-relativistic spacetimes. Therefore, following the studies of
the holographic ferromagnetic-paramagnetic phase transition in the
presence of nonlinear electrodynamics \cite{Zhang2plb}, here we
would like to extend the investigation on this system in the
background of Lifshitz spacetime by taking into account the
nonlinear PM electrodynamics. The motivation for taking into
account the PM electrodynamics instead of Maxwell one comes from
the fact that the Maxwell field is conformally invariant, and
hence the corresponding energy momentum tensor is traceless, only
in four dimensions. A natural question then arises: Is there an
extension of Maxwell action in arbitrary dimensions that is
traceless and hence possesses the conformal invariance? The answer
is positive and the conformally invariant Maxwell action in
$d$-dimensional spacetime is given by \cite{shamsip30},
\begin{equation} \label{act2}
I_{PM}=\int{d^{d}x \sqrt{-g}\left(-\mathcal{F}\right)^q},
\end{equation}
where $\mathcal{F}=F_{\mu\nu}F^{\mu\nu}$ is the Maxwell invariant
and $q$ is the power parameter. One can easily check that action
(\ref{act2}) is invariant under conformal transformation
$g_{\mu\nu}\to\Omega^{2}g_{\mu\nu}$ and $A_{\mu}\to A_{\mu}$. The
associated energy-momentum tensor of the above action is given by
\begin{equation}
T_{\mu\nu}=2\left(q
F_{\mu\rho}F{^{\rho}_{\nu}}F^{q-1}-\frac{1}{4}g_{\mu\nu}\mathcal{F}^{q}\right).
\end{equation}
It is easy to check that the above energy-momentum tensor is
traceless for $q={d}/{4}$. The theory of conformally invariant
Maxwell field is considerably richer than that of the linear
standard Maxwell field and in the special case $(q=1)$ it recovers
the Maxwell action \cite{SheyPM,SheyPM2,SheyPM3}. It is worthwhile
to investigate the effects of exponent $q$ on the
paramagnetism-ferromagnetism phase transition in the Lifshitz
background. To be more general, in this work, we consider not only
the conformal case where $d=4p$, but also the arbitrary value of
$q$. This allows us to consider more solutions from different
perspective \cite{dehyadegari} and brings rich physics in studying
holographic paramagnetism-ferromagnetism phase transition. This
holographic model can provide a powerful tool to analyze phenomena
involving magnetization. In particular, we shall investigate how
the PM electrodynamics influences the critical temperature and
magnetic moment. Interestingly, we find that the effect of
sub-linear PM field can lead to the easier formation of the
magnetic moment at higher critical temperature with respect to
other kinds of nonlinear electrodynamics. In other words, the
critical temperature increases with decreasing the value of the
power parameter. For higher values of the power parameter, the gap
in the magnetic moment in the absence of magnetic field, is
smaller which in turn exhibits that the condensation is formed
harder.

We shall focus on four- and five-dimensional holographic
paramagnetic-ferromagnetic phase transition in probe limit by
neglecting the back reaction of both gauge and the $2-$form fields
on the background geometry. We employ the numerical shooting
method to investigate the features of our holographic model. All
theses holographic ferromagnetic models are constructed only in
the relativistic spacetimes. We wonder whether this model still
hold in non-relativistic spacetimes for example the Lifshitz
spacetime, which is our motivation in this paper.

This paper is organized as follows; In section \ref{setup}, we
introduce the action and basic field equations of the holographic
model for paramagnetic-ferromagnetic phase transition in the
Lifshitz black hole with PM electrodynamics. In section
\ref{numst}, we employ the shooting method for our numerical
calculations and obtain the critical temperature and magnetic
moment. We also study the magnetic susceptibility density. In the
last section, we summarize our results and concluding remarks.
\section{The Holographic Set-up}\label{setup}
In many condensed matter systems, it can be found that the phase
transitions governed by fixed points which exhibit the anisotropic
scaling of spacetime $t\to b^{z}t$, $x\to b x$, where
$z$ is the Lifshitz dynamical exponent. Let us consider the
gravity description dual for this scaling in the $d-$
dimensional spacetime of the form
\begin{equation}
ds^{2}=-r^{2z}f(r)dt^{2}+\frac{dr^{2}}{r^{2}f(r)}+r^{2}\sum_{i=1}^{d-2}dx_{i}^{2},
\label{1s}
\end{equation}%
with
\begin{equation}
 f(r)=1-\left(\frac{r_{+}}{r}\right)^{z+d-2},
\end{equation}%
where $r_{+}$ is the event horizon radius of the black hole. The
Hawking temperature of black hole, on the horizon, which can be
interpreted as the temperature of CFT, is given by \cite{q.pan}
\begin{equation}
T=\frac{f^{\prime }(r_{+})}{4\pi }=\frac{(z+d-2){r_{+}^{z}}}{4\pi }.
\label{3}
\end{equation}%

The critical points with $z>1$ are often called to be non-relativistic.
The Lifshitz spacetime in a $d-$dimensional background can be realized by the following action
\begin{eqnarray}
S&=&\frac{1}{2\kappa ^{2}}\int
d^{d}x\sqrt{-g}\left(R-{2}{\Lambda}+ L_{1}\left(
\mathcal{F}\right) +\lambda^{2} L_{2}\right),\label{Act}
\end{eqnarray}%
where $\kappa ^{2}=8\pi G$ with $G$ is Newtonian gravitational
constant, $g$ is the determinant of metric, $R$ is Ricci scalar
and $\Lambda=-{(z+d-3)(z+d-2)}/{2l^{2}}$ is the
cosmological constant of $d-$dimensional AdS spacetime with radius
$l$. The action of Power-Maxwell filed is taken as power-law
function of the form  $L_{1}(\mathcal{F})=-\beta
\mathcal{F}^{q}$, where $\beta$ is a constant, $q$ is the power
parameter of the Power-Maxwell field \cite{PM1,shamsip30}. Here
$\mathcal{F}=F_{\mu \nu }F^{\mu \nu }$ is the Maxwell invariant in
which $F_{\mu \nu}=\nabla _{\lbrack \mu }A_{\mu ]}$ and $A_{\mu}$
is the gauge potential of U(1) gauge field. Clearly $q=1$
corresponds to the Maxwell electrodynamics,
$L_{1}=-\mathcal{F}/4$, and the Einstein-Maxwell theory is
recovered. In action (\ref{Act}) the Lagrangian density $L_{2}$
consists a $U(1)$ field $A_{\mu}$ and a massive 2-form field
$M_{\mu\nu}$ in $(d)$-dimensional spacetime and is given by
\cite{Cai3}
\begin{eqnarray}{\nonumber}
L_{2}=-\frac{1}{12}(dM)^{2}-\frac{m^{2}}{4}M_{\mu\nu}M^{\mu\nu}-\frac{1}{2}M^{\mu\nu}F_{\mu\nu}
-\frac{J}{8}V(M),
  \label{Act1}
\end{eqnarray}
where $dM$ is the exterior differential of $2$-form field
$M_{\mu\nu}$, $\lambda$ and $J$ are two real parameters with $J<0$
for producing the spontaneous magnetization and $\lambda^{2}$
characterizes the back reaction of two polarization field
$M_{\mu\nu}$, and the Maxwell field strength on the background
geometry. In addition, $m$ is the mass of $2$-form field
$M_{\mu\nu}$ being greater than zero \cite{Cai3} and $dM$ is the
exterior differential $2$-form field $M_{\mu\nu}$. $V(M_{\mu\nu})$
is a nonlinear potential of $2$-form field $M_{\mu\nu}$ describing
the self-interaction of polarization tensor which should be
expanded as the even power of $M_{\mu\nu}$. In this model, for
simplicity, we take the following form for the potential
 \begin{equation}
 V\left(M\right) =\left( ^{*}M_{\mu\nu}{M^{\mu\nu}}\right)^{2}=[^{*}(M \wedge M)]^{2},
 \label{potential}
\end{equation}%
where $*$ is the Hodge star operator \cite{Cai3}.

Varying action (\ref{Act}), we can get the equations of motion for
the matter fields as
\begin{eqnarray}
 0&=& \nabla ^{\tau }(dM)_{\tau\mu\nu}-\notag \\
 &&-m^{2}M_{\mu\nu}-J (^{*}M_{\tau\sigma}M^{\tau\sigma})(^{*}M_{\mu\nu})-F_{\mu\nu} \,  \label{01} \\
 0&=& \nabla ^{\mu }\left(q{F_{\mu\nu}}{(\mathcal{F})^{q-1}}+\frac{\lambda^{2}}{4}M_{\mu\nu}\right).  \label{02}
\end{eqnarray}
In the probe limit, we can neglect the back reaction of the 2-form
and PM fields on the background Lifshitz geometry (\ref{1s}). In
order to explore the effects of Lifshitz dynamical exponent
$z$ and power parameter $q$ on the holographic ferromagnetic
phase transition, we take the self-consistent ansatz with matter
fields based on the Lifshitz spacetime as follows,
\begin{equation}
M_{\mu\nu}=-p(r)dt{\wedge}dr+ \rho(r)dx{\wedge}dy ,  \label{M}
\end{equation}%
\begin{equation}
A_{\mu}=\phi(r)dt+ B xdy,  \label{A}
\end{equation}%
where $B$ is a uniform magnetic field which is considered as an
external magnetic field of dual boundary field theory.
$\rho(r)$, $p(r)$ and $\phi(r)$ which are two components of the
polarization field and the electrical potential,respectively, are function
of $r$. Inserting this ansatz into Eqs. (\ref{01}) and (\ref{02}), nontrivial
equations of motion in $d$-dimensional Lifshitz spacetime can be
obtained as follows,
\begin{eqnarray}
0 &=&\rho ^{\prime \prime }+\rho ^{\prime }\left[ \frac{%
    f^{\prime }}{f}+\frac{(z+1)-(6-d)}{r}\right] -\notag\\
&&-\frac{\rho
}{r^{2}f}\left[m^{2}+\frac{4Jp^{2}}{r^{2z-2}}\right]+\frac{B}{r^{2}f} ,
\notag \\
0 &=&\left(m^{2}-\frac{4J\rho^{2}}{r^4}\right)p-\phi ^{\prime}, \notag \\
0&=& \phi ^{\prime \prime}+\frac{\phi ^{\prime}} {r((2q-1)\phi^ {\prime 2}-B^{2}r^{2z-6})}
\left[-B^{2}r^{2z-6}(z-\right.
    \nonumber\\&& \left.-(d+3)+4 q)+\phi ^{\prime 2}\left(2q(z-1)-[z+(6-d)]\right)
    \right]\nonumber\\&& +\frac{\lambda ^{2}}
    {2^{q+1}r^{2z+4}}\left(p^{\prime}+\frac{d-z-1}{r}p\right)\notag\\
    &&\times \left[\frac{ (\phi ^{\prime 2}-B^{2}r^{2z-6}) ^{2-
            q}} {(2 q -1)\phi ^{\prime 2}-{B^{2}}{r^{2z-6}}}\right] ,
\,\label{EOM}
\end{eqnarray}%
where the prime stands for the derivative with respect to $r$. In
the limiting case where $d=4$, $ q=1$ and $z=1$,
the above equations reduce to the standard holographic
paramagnetism-ferromagnetism phase transition models discussed in
Ref.~\cite{Cai3}. In order to solve Eq. (\ref{EOM}) numerically,
we need to seek the boundary conditions for $\rho, \phi$ and $p$
near the black hole horizon at the spatial infinity. Therefore, in
additional to $f(r_{+})=0$,  and due to the fact that the norm of
the gauge field namely $g_{\mu\nu}A^{\mu}A^{\nu}$ should be finite
at the horizon, it is required $\phi(r_{+})=0$ by considering the
regularity condition for $\rho(r_{+})$ at the horizon. The
asymptotic solutions for matter fields governed by the field
equations (\ref{EOM}) near the boundary ($r\rightarrow \infty $)
are given by
\begin{gather}
\phi (r)\sim \mu
-\frac{\sigma^{\frac{1}{2q-1}}}{r^{\frac{((d-3)+2q)+z(1-2q)}{2q-1}-1}},\notag\\
p(r)\sim \frac{\sigma^{\frac{1}{2q-1}} \left[\frac{((d-3)+2q)+z(1-2q)}{{2q-1}}-1\right]}{m^{2}r^{\frac{((d-3)+2q)+z(1-2q)}{2q-1}}},\notag\\
\rho (r)\sim \frac{\rho _{-}}{%
    r^{\Delta _{-}}}+\frac{\rho _{+}}{r^{\Delta_{+}}}+\frac{B}{m^{2}}, \label{boundval}
\end{gather}%
where $\mu $ and $\sigma $ are respectively interpreted as the
chemical potential and charge density in the dual field theory, and
\begin{equation}
\Delta _{\pm
}=\frac{(6-d)-z}{2}\pm \frac{1}{2}\sqrt{4m^{2}-[z-(6-d)]^{2}}.
\end{equation}
The coefficients $\rho_{+}$ and $\rho_{-}$ are two constants
correspond to the source and vacuum expectation value of dual
operator in the boundary field theory when $B=0$. Therefore,
condensation happens spontaneously below a critical temperature
when we set $\rho_{+}=0$. Considering $B\ne0$, the asymptotic
behavior is governed by external magnetic field $B$. It is
important to note that, unlike other nonlinear electrodynamics
such as Born-Infeld-like electrodynamics, the boundary condition
for the gauge field $\phi$ depends on the power parameter $q$ of
the PM field and the dynamical exponent,$z$,  \cite{Zhao, Jing}.
Using the asymptotic behaviour given by Eq. (\ref{boundval}) and
the fact that $\phi$ should be finite as $r\to\infty$, it leads to
\begin{equation}
\frac{(d-3)+2q+z(1-2q)}{2q-1}-1>0,  \Rightarrow
1/2<q<\frac{d-2+z}{2z}.
\end{equation}
Thus, the power parameter $q$ cannot take an arbitrary value and
is bounded by the Lifshitz exponent $z$ and spacetime dimensions
$d$. Besides, the asymptotic behavior of $p(r)$ confined by these
parameters. Basically $q$ should satisfy the above condition,
e.g., for $z=3/2$ in four dimensions ($d=4$) we have $q<7/6$,
while in five dimensions ($d=5$), the upper bound for power
parameter becomes $q<9/6$.

In the remaining part of this paper, we will study the holographic
ferromagnetic-paramagnetic phase transition numerically.
  \begin{figure*}
    \centering{

    \subfigure[~$\protect$$z=3/2$]{
        \label{fig1a}\includegraphics[width=.35\textwidth]{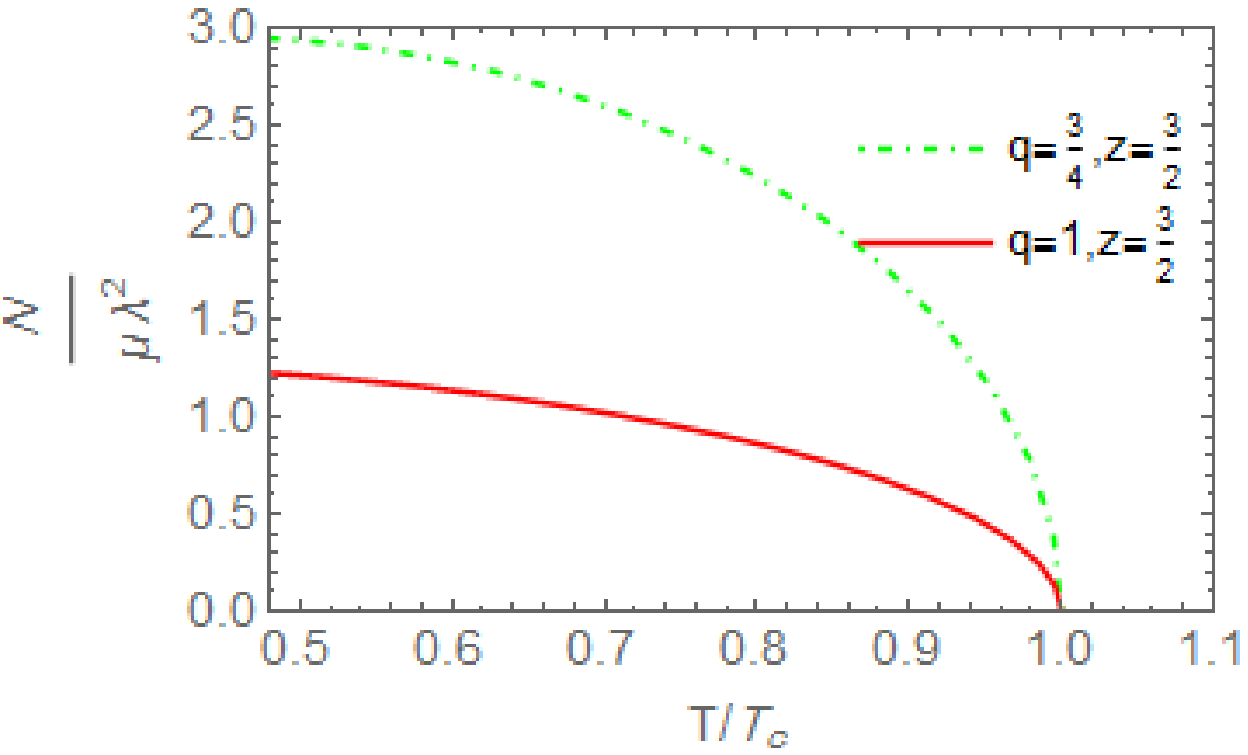}\qquad}
    \subfigure[$z=7/4$]{
        \label{fig1b}\includegraphics[width=.35\textwidth]{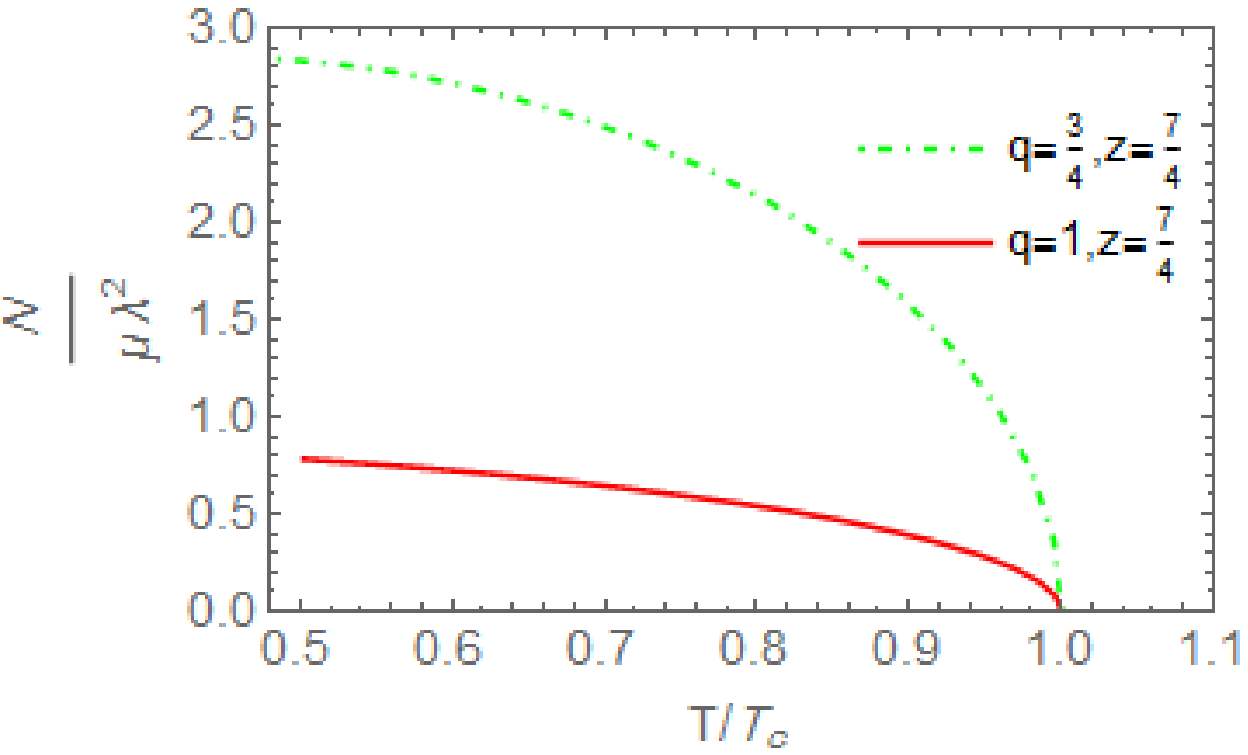}\qquad}}

    \caption{The behavior of magnetic moment $N$ and the critical
        temperature with different values of power parameter $q$, in different dynamical exponents $z$ for $m^{2}=1/8$ and $J=-1/8$ for $d=4$.}
    \label{fig1}
  \end{figure*}
  \begin{table*}
    \centering%
    \caption{Numerical results of ${T_{c}}/{\protect\mu }$ for
        different values of $\protect q $ in different dynamical exponents $z$ for $d=4$.}
    \begin{tabular}{llllll}
        \hline
        $d=4$&$ z=3/2 $ & $z=7/4 $ \\
        \hline
        $q=3/4$&  $2.7245$& $2.2219$     \\
        $q=4/4$&  $0.8584$& $0.4490$     \\

        \hline
    \end{tabular}
    \label{Table1}
      \end{table*}

  \begin{table*}
    \centering%
    \caption{The magnetic moment $N$ with different values of $q$ and $z$ for $d=4$.}
    \begin{tabular}{llllll}
        \hline
        $z$&$ 3/2 $& $7/4 $    \\
        \hline
        $q=3/4$ & $8.5730(1-T/Tc)^{1/2} $& $8.3168(1-T/Tc)^{1/2}$   \\
        $q=1$&  $ 3.6318(1-T/Tc)^{1/2} $& $2.3252(1-T/Tc)^{1/2}$   \\

        \hline
    \end{tabular}
    \label{Table2}

  \end{table*}

  \begin{figure*}
    \centering{
            \subfigure[~$\protect$$z=3/2$]{
        \label{fig2a}\includegraphics[width=.35\textwidth]{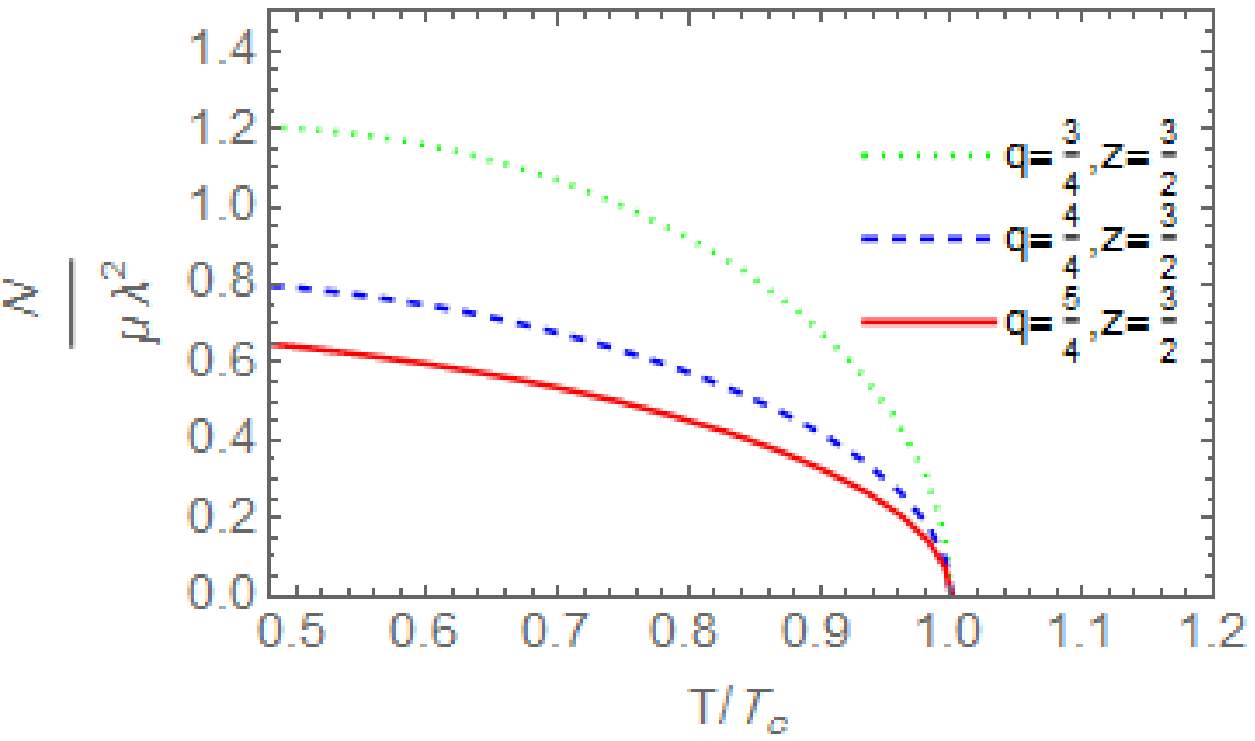}\qquad}
    \subfigure[~$\protect$$z=17/10$]{
        \label{fig2b}\includegraphics[width=.35\textwidth]{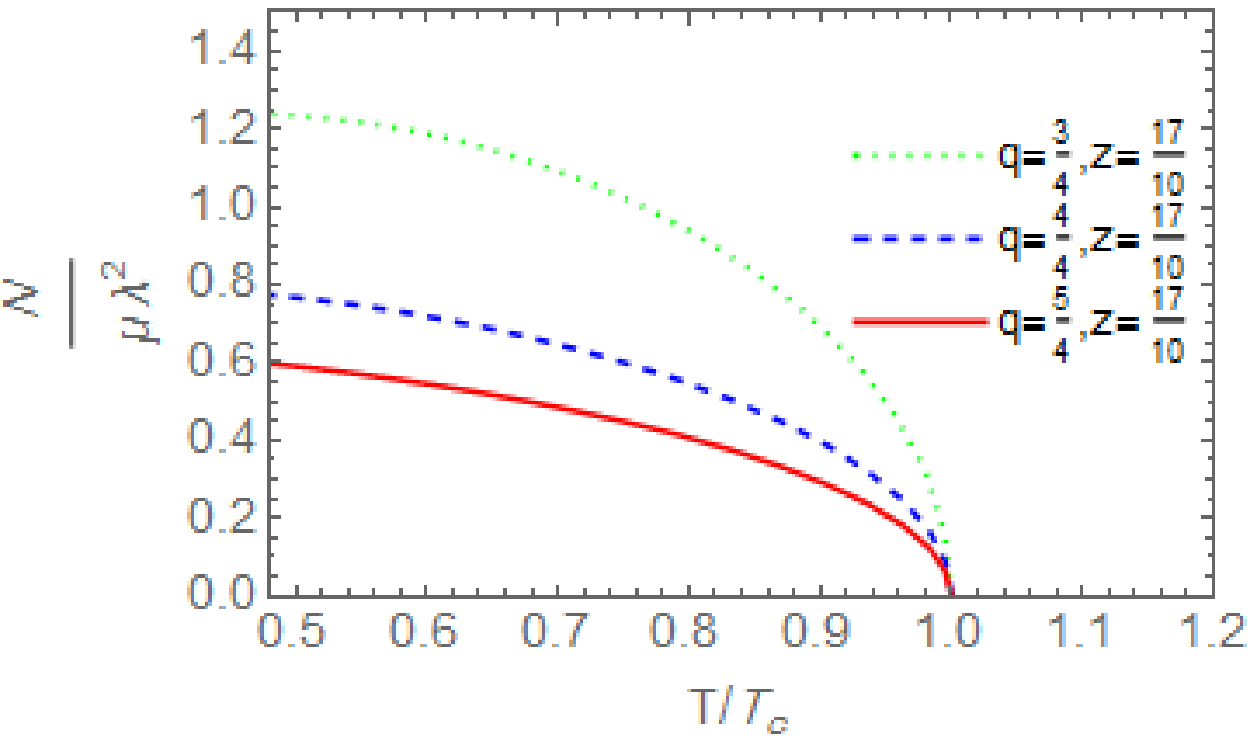}\qquad}}

    \caption{The behavior of magnetic moment $N$ and the critical
        temperature with different values of power parameter $q$, in different dynamical exponents $z$ for $m^{2}=1/8$ and $J=-1/8$ for $d=5$.}
    \label{fig2}
  \end{figure*}

\begin{table*}[!]
    \centering%
    \caption{Numerical results of ${T_{c}}/{\protect\mu }$ for
        different values of $\protect q $ in different dynamical exponents $z$ for $d=5$.}
    \begin{tabular}{llllll}
        \hline
        $d=5$&  $z=3/2$& $z=17/10$ \\
        \hline
        $q=3/4$&  $2.8317$& $2.5128$  \\
        $q=4/4$&  $1.5054$& $1.2261$ \\
        $q=5/4$&  $1.1286$& $0.8736$  \\
        \hline
    \end{tabular}

 \label{Table3}

 \end{table*}

 \begin{table*}[!]
 \centering%
  \caption{The magnetic moment $N$ with different values of $q$ and $z$ for $d=5$.}
 \begin{tabular}{llllll}
 \hline
 $z$&$ 3/2 $& $17/10 $    \\
 \hline
 $q=3/4$ & $3.4779(1-T/Tc)^{1/2} $& $3.5974(1-T/Tc)^{1/2}$   \\
 $q=1$&  $ 2.3363(1-T/Tc)^{1/2} $& $2.2563(1-T/Tc)^{1/2}$   \\
 $q=5/4$& $ 1.8779(1-T/Tc)^{1/2} $& $1.7178(1-T/Tc)^{1/2}$   \\
 \hline
 \end{tabular}
 \label{Table4}

 \end{table*}
\section{Numerical calculation for spontaneous magnetization and susceptibility\label{numst}}
In the previous section, we have investigated the holographic
paramagnetic-ferromagnetic phase transition in arbitrary
dimensions and derived all expressions in terms of spacetime
dimension $d$. However, in the numerical calculations one needs to
specify the spacetime dimensions $d$. Thus, in this section we
focus on 4D and 5D in our numerical calculations which are more
realistic from physical point of view. We set our calculations in
the grand canonical ensemble where the chemical potential $\mu$ is
fixed. The expression of magnetic moment in the background of
Lifshitz black hole will be changed as
\begin{equation}
 N=-\lambda^{2}\int\frac{\rho}{2 r^{d-z-1}}dr.
 \end{equation}
 We can obtain the basic features of the model by choosing $m^{2}=-J=1/8$ and
$\lambda=1/2$ as a typical example in the numerical computation.
This is due to the fact that the choice of parameters will not
qualitatively modify our results. Using the shooting method
\cite{hartnoll}, we can solve Eq.(\ref{EOM}) numerically to
investigate the holographic phase transition, and discuss the
effects of PM electrodynamics and the dynamical exponent $z$
on the magnetic moment in different dimensions. Hereafter, we define the dimensionless
coordinate $Z=r_{+}/r$ instead of $r$, since it is easier to work
with it. The numerical calculation may be justified by virtue of
the field equation symmetry,
\begin{equation*}
r\rightarrow ar,\text{ \ \ } t\rightarrow a^{-z}t,\text{ \ \
}f\rightarrow a^{2}f,\text{ \ \ } \phi \rightarrow a\phi ,\text{ \
\ }\mu \rightarrow
 a^{z}\mu.
\end{equation*}%
We can use the above scaling symmetry, and obtain the solution of
Eqs. (\ref{EOM}) with the same chemical potential as it is
discussed in Ref.~\cite{binaei} in details. We consider the cases
of different power parameter $q$ in $4D$ and $5D$ spacetime with
different allowed values of Lifshitz parameter. We present our
results in Figs.\ref{fig1} and \ref{fig2}. In these figures, we
plot the magnetic moment for $d=4$ and $d=5$, for allowed values
of $z$, with different values of power parameters. When the
temperature is lower than $T_{C}$, we find that in the absence of
an external magnetic field, the spontaneous condensate of $\rho$
(magnetic moment) in the bulk appears. In the vicinity of critical
temperature, the numerical results show that the second order
phase transition happens which its behavior can be obtained by
fitting this curve ($N\propto\sqrt{1-T/T_{C}}$). The results have
been shown in Tables \ref{Table2} and \ref{Table4} . We find that
there is a square root behavior for the magnetic moment versus
temperature, and the critical exponent is the same as that of mean
field theory for these two dimensions. In other
 words, the holographic paramagnetism-ferromagnetism phase transition still
exists in Lifshitz black hole spacetime in the presence of PM
electrodynamics, similar to the cases of Born-Infeld-like
nonlinear electrodynamics \cite{Zhang2plb}.
 \begin{figure*}
    \centering{
    \subfigure[~$\protect$$z=3/2, q=3/4$]{
        \label{fig3a}\includegraphics[width=.45\textwidth]{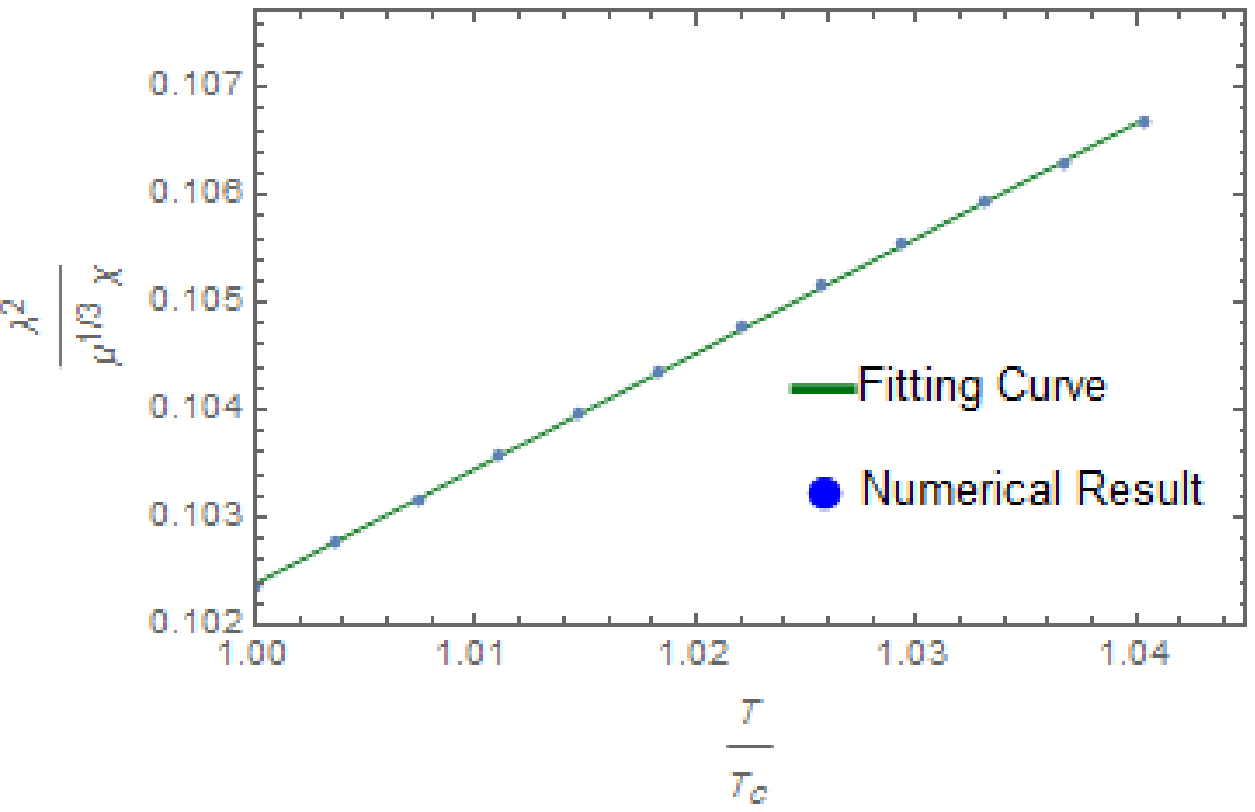}\qquad}
    \subfigure[~$\protect$$z=7/4, q=3/4$]{
        \label{fig3b}\includegraphics[width=.45\textwidth]{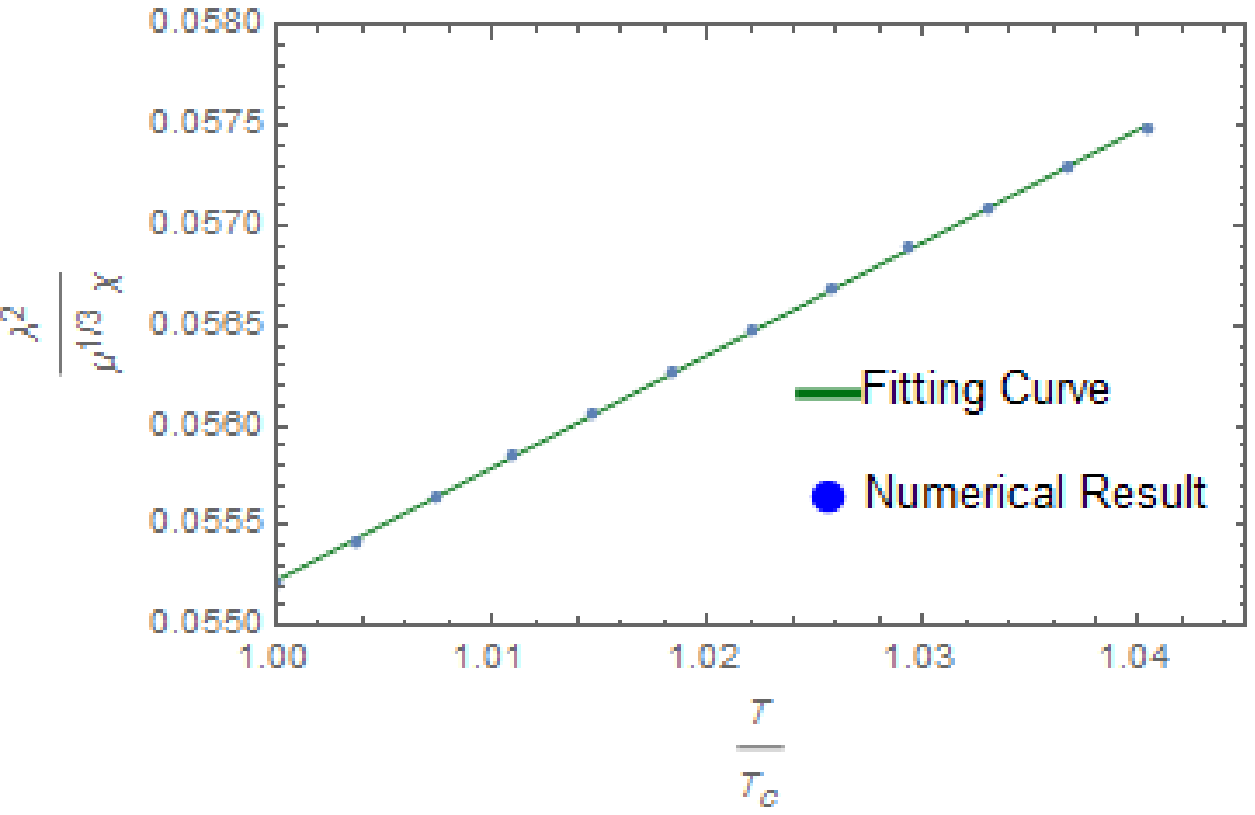}\qquad}
    \subfigure[~$\protect$$z=3/2, q=1$]{
        \label{fig3c}\includegraphics[width=.45\textwidth]{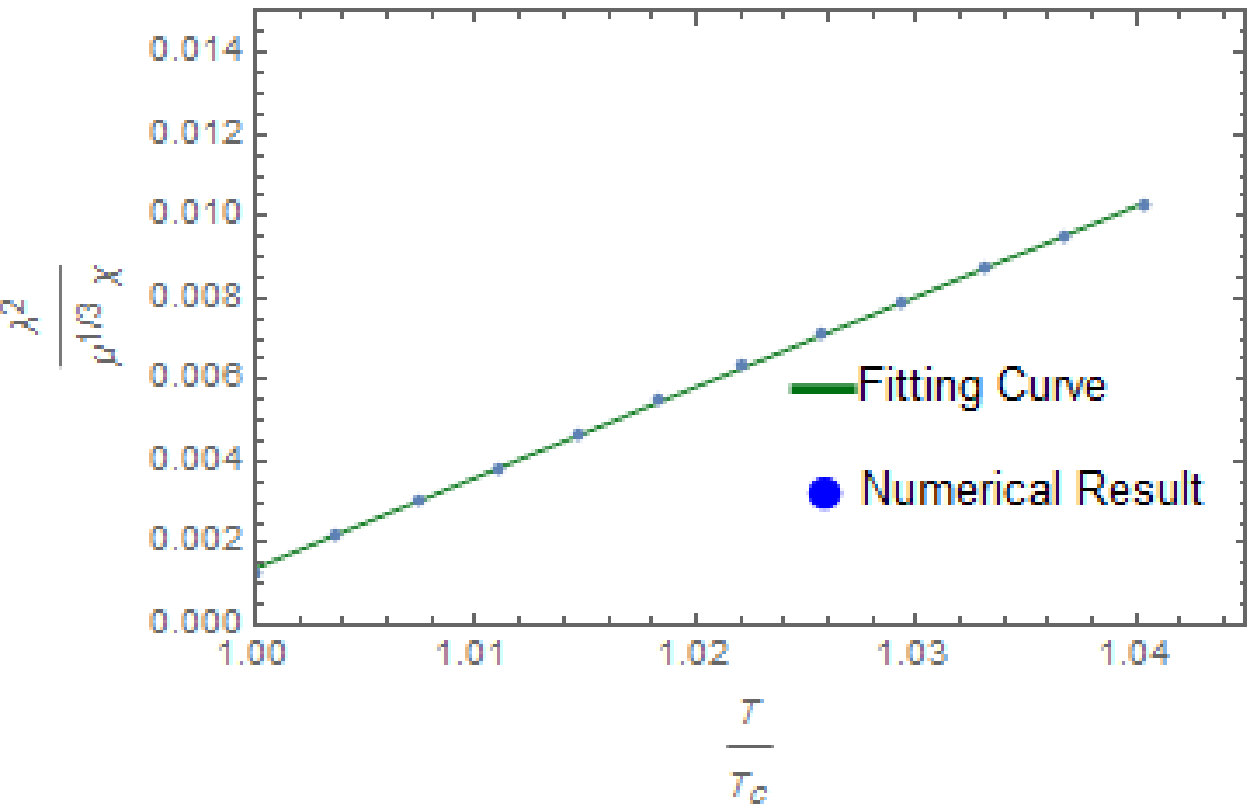}\qquad}
    \subfigure[~$\protect$$z=7/4, q=1$]{
        \label{fig3d}\includegraphics[width=.45\textwidth]{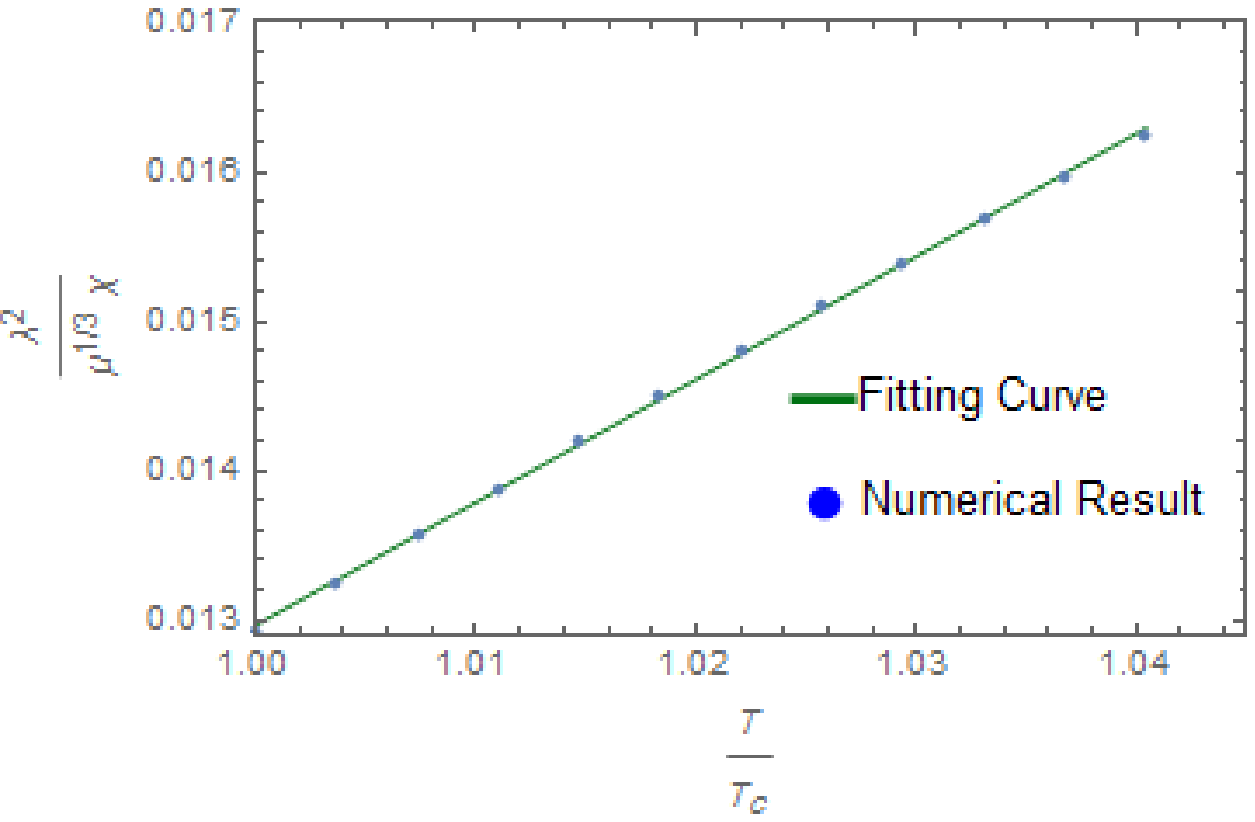}\qquad}}

    \caption{The behavior of inverse susceptibility density as a function of temperature
        with different values of $q$, at different dynamical exponent $z$ for $d=4$.}
    \label{fig3}
 \end{figure*}
\begin{table*}
    \centering%
    \caption{The magnetic susceptibility ${\lambda^{2}/\chi\mu^{1/3}}$ with different values
        of $z$ and $q$.}
    \begin{tabular}{llllll}
        \hline
        $d=4$& $$  & $z=3/2 $ & $z=7/4$   \\
        \hline
        $q=3/4$& $\lambda^{2}/\chi\mu^{1/3}$ & $0.1071(T/Tc-22.4235)$  & $0.0564(T/Tc-49.5533)$    \\
        $$& $\theta/\mu$ & $-0.1215$  & $-0.477$    \\
        $q=1$& $\lambda^{2}/\chi\mu^{1/3}$ & $0.2215(T/Tc-1.0063)$  & $0.0823(T/Tc-1.1869)$    \\
        $$& $\theta/\mu$ & $-0.8530$  & $-0.3782$    \\

        \hline

        \end{tabular}

        \label{table5}
        \end{table*}

 \begin{figure*}[h]

    \centering{
    \subfigure[~$\protect$ $z=3/2$, $q=3/4$]{
        \label{fig4a}\includegraphics[width=.45\textwidth]{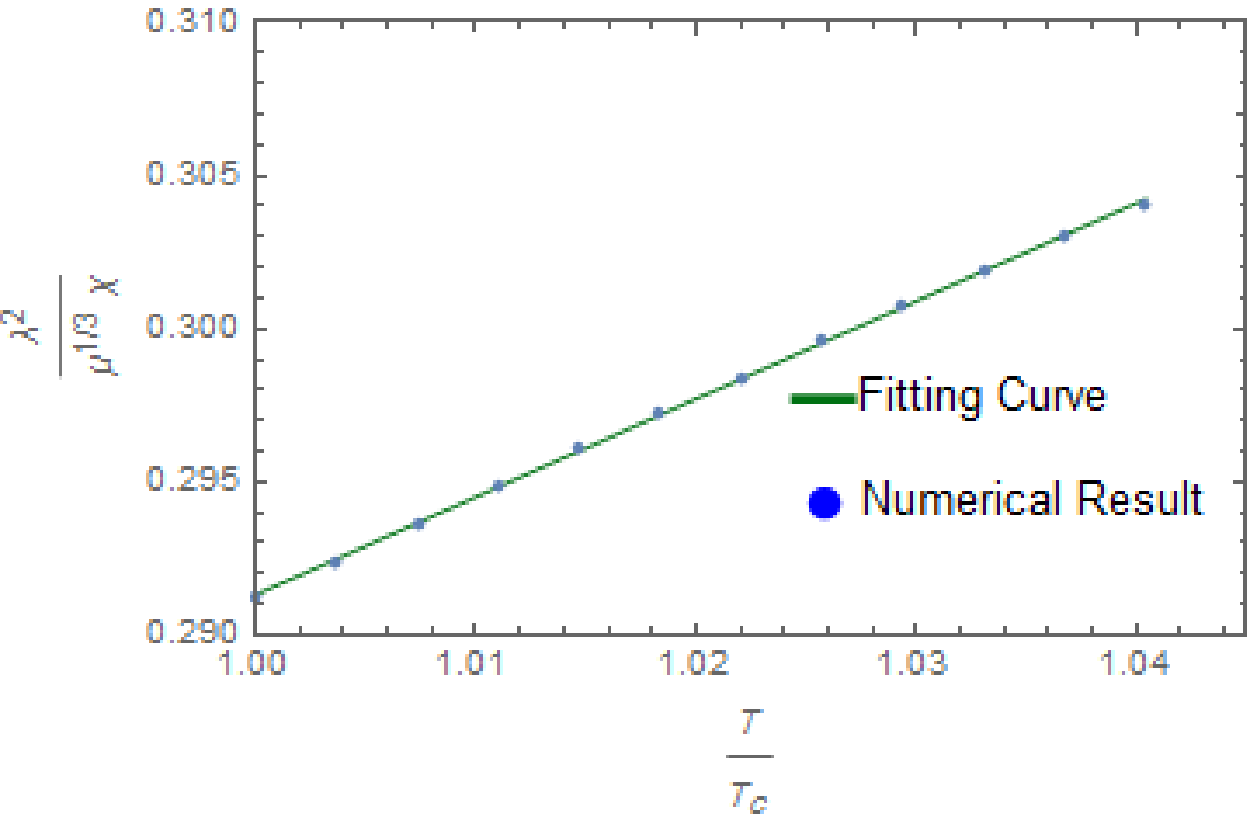}\qquad}
    \subfigure[~$\protect$ $z=17/10$, $q=3/4$]{
        \label{fig4b}\includegraphics[width=.45\textwidth]{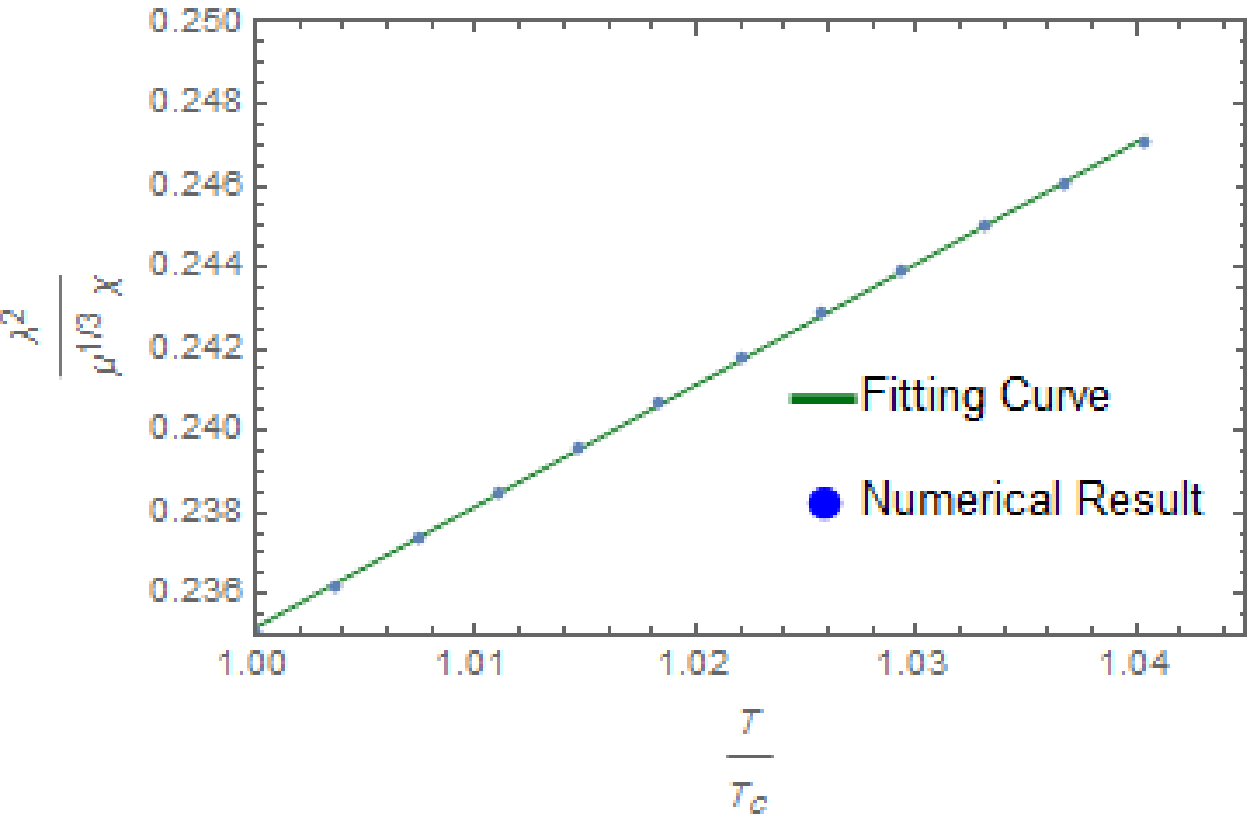}\qquad}
    \subfigure[~$\protect$ $z=3/2$, $q=1$]{
        \label{fig4c}\includegraphics[width=.45\textwidth]{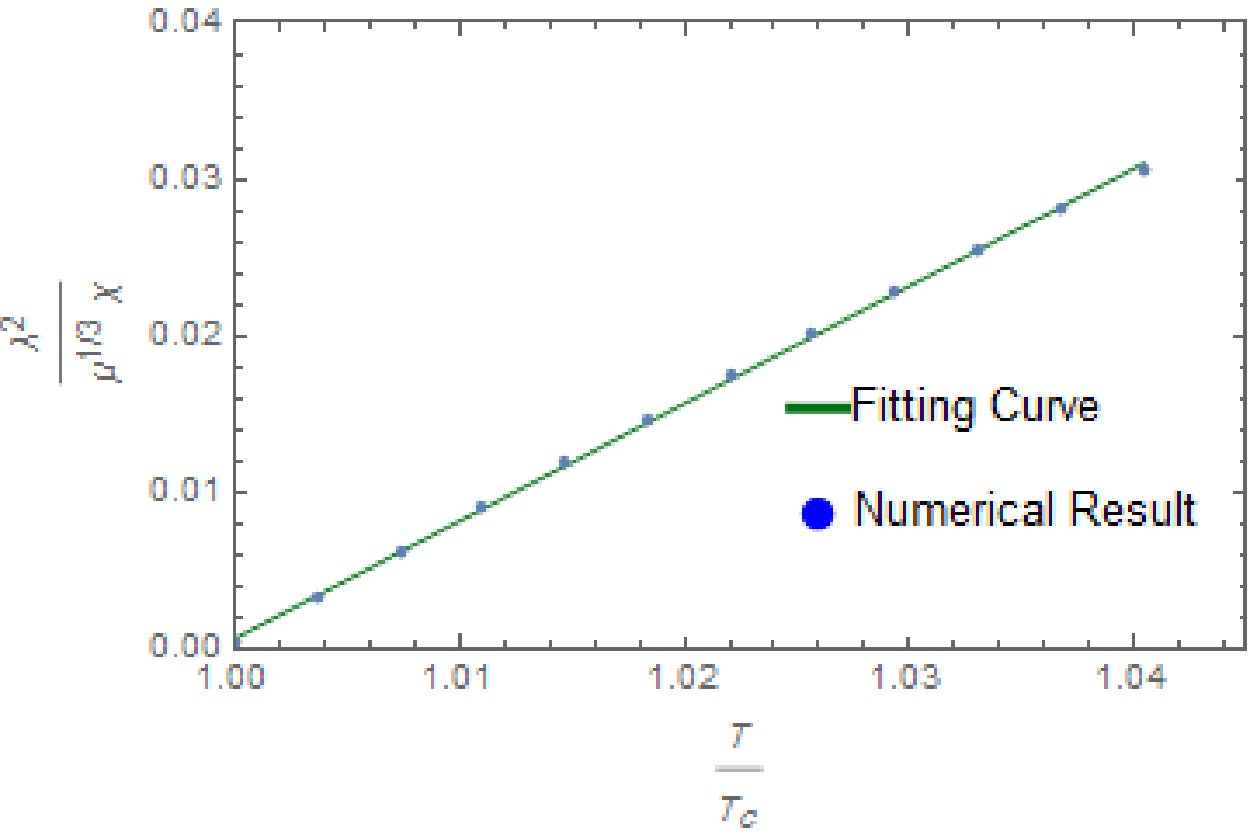}\qquad}
    \subfigure[~$\protect$ $z=17/10$, $q=1$]{
        \label{fig4d}\includegraphics[width=.45\textwidth]{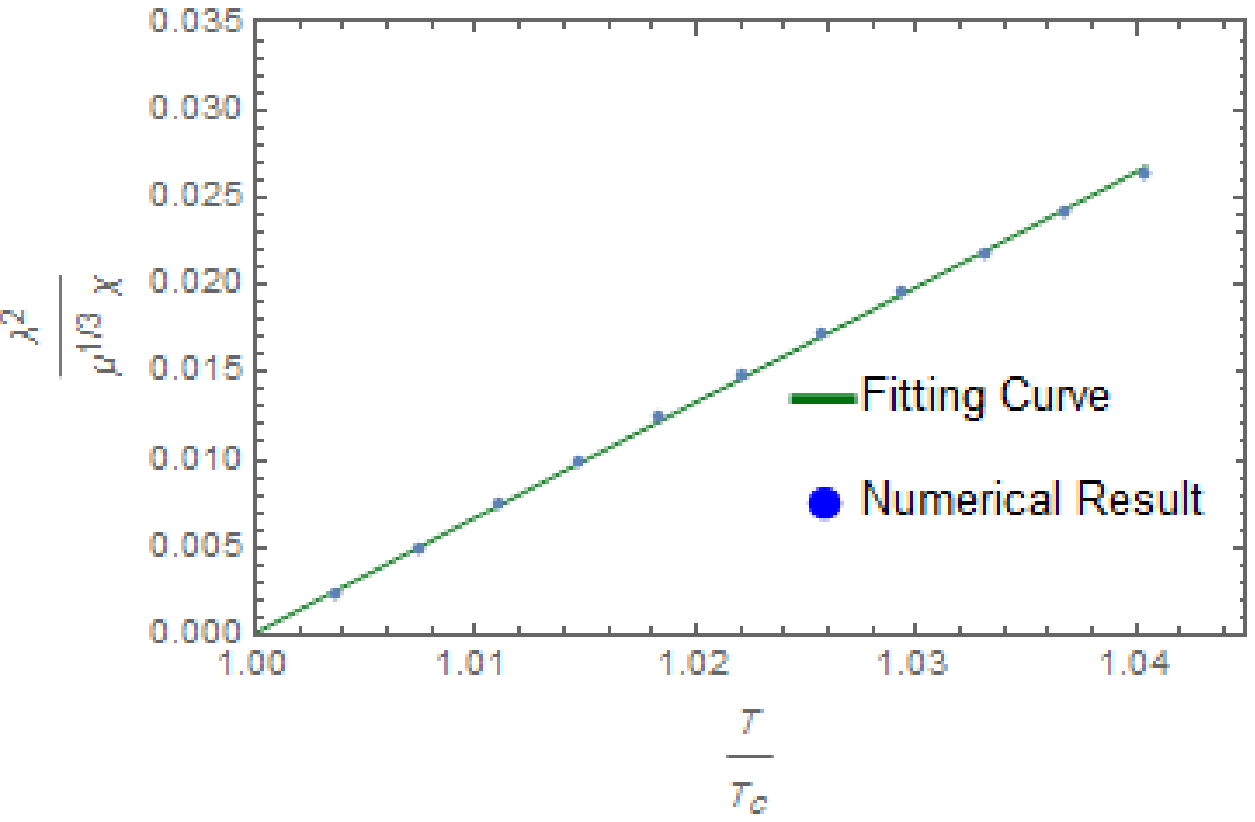}\qquad}
    \subfigure[~$\protect$ $z=3/2$, $q=5/4$]{
        \label{fig4f}\includegraphics[width=.45\textwidth]{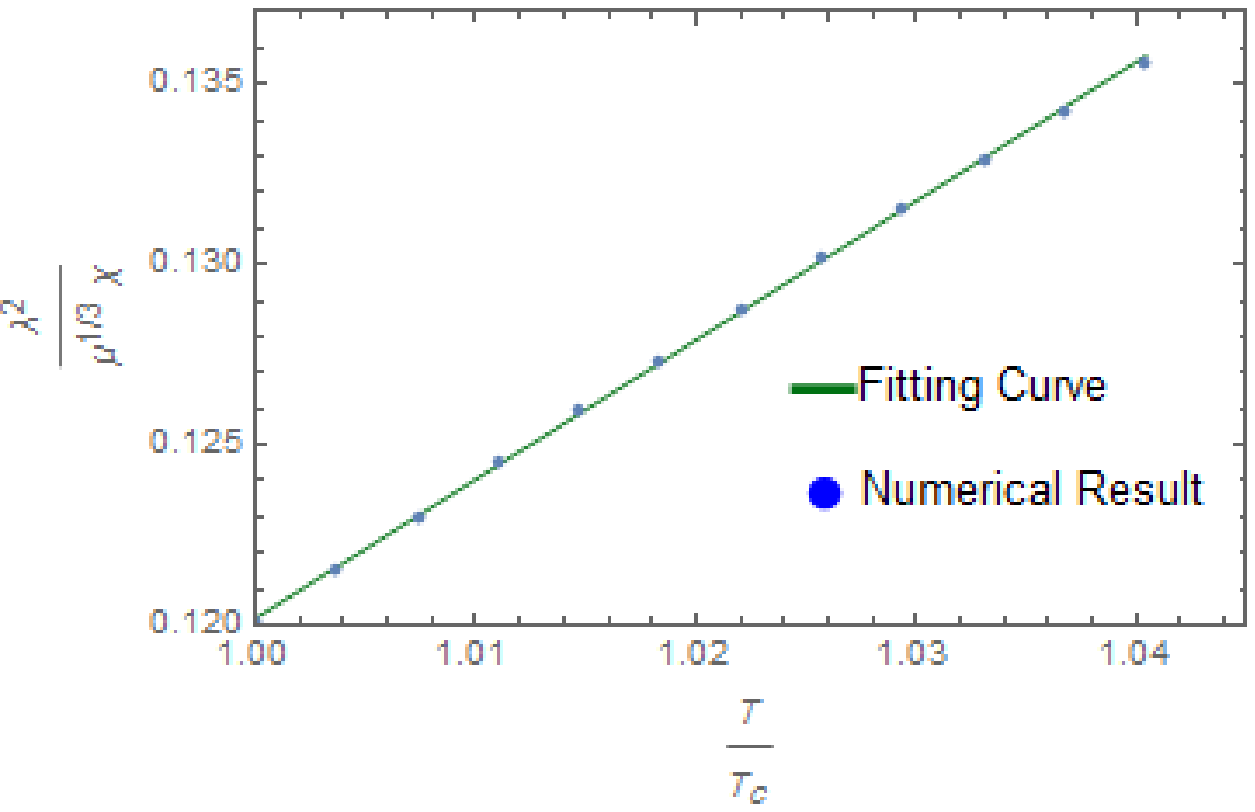}\qquad}
    \subfigure[~$\protect$ $z=17/10$, $q=5/4$]{
        \label{fig4e}\includegraphics[width=.45\textwidth]{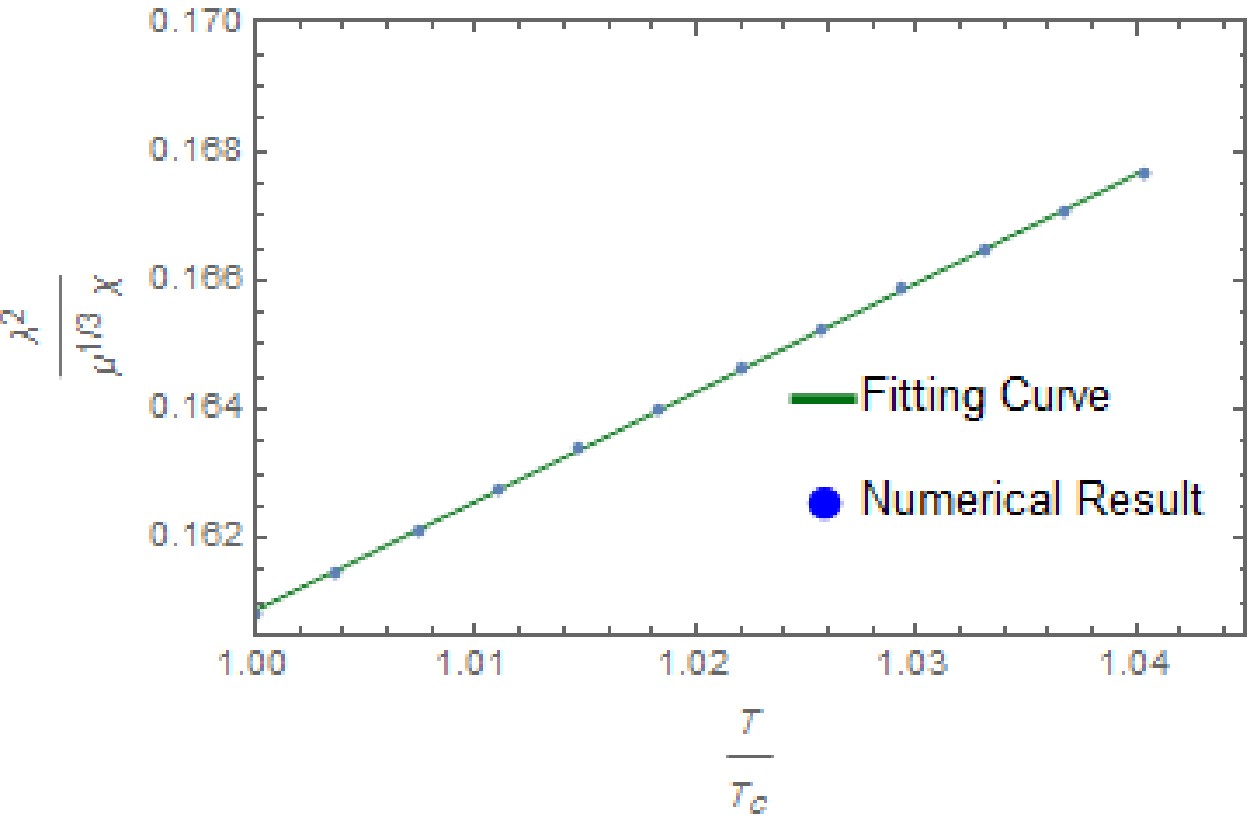}\qquad}}

    \caption{The behavior of inverse susceptibility density as a function of temperature
        with different values of $q$, at different dynamical exponent $z$ for $d=5$.}
    \label{fig4}
 \end{figure*}

\begin{table*}[!]
    \centering%
    \caption{The magnetic susceptibility ${\lambda^{2}/\chi\mu^{1/3}}$ with different values
        of $z$ and $q$.}
    \begin{tabular}{llllll}
        \hline
        $d=5$& $$  & $z=3/2 $ & $z=17/10$   \\
        \hline
        $q=3/4$& $\lambda^{2}/\chi\mu^{1/3}$ & $0.3198(T/Tc-11.2191)$  & $0.2968(T/Tc-4.8108)$    \\
        $$& $\theta/\mu$ & $-0.2524$  & $-0.5223$    \\
        $q=1$& $\lambda^{2}/\chi\mu^{1/3}$ & $0.7515(T/Tc-1.0008)$  & $0.6584(T/Tc-1.0002)$    \\
        $$& $\theta/\mu$ & $-1.5042$  & $-1.2258$    \\
        $q=5/4$& $\lambda^{2}/\chi\mu^{1/3}$ & $0.3861(T/Tc-1.4518)$  & $0.1692(T/Tc-20.3897)$    \\
        $$& $\theta/\mu$ & $-0.7773$  & $-0.0428$    \\
        \hline

    \end{tabular}

    \label{table6}
\end{table*}

From Figs.\ref{fig1} and\ref{fig2}, we observe that with increasing the
value of power parameter $q$, the magnetic moment decreases for
the fixed values of $z$ in different dimensions. Similar behavior is seen for the
cases of Born-Infeld-like nonlinear electrodynamics. This implies
that the magnetic moment is harder to form in the presence of
nonlinear electrodynamics. This is consistent with the results
given in Ref. \cite{Zhang2,Wu1}. This behavior has been also seen for the
holographic superconductor in the Schwarzschild- AdS black hole,
where the three types of nonlinear electrodynamics make scalar
condensation harder to be formed \cite{Zhao}.

In Tables \ref{Table1} and \ref{Table3} our numerical results for
critical temperature with different values of $q$ and $z$ are
presented. For $d=4$, when $z\rightarrow 1$, in the Maxwell limit
($q=1$), our numerical results reproduce the results of Ref.
\cite{Zhang2}. We see from Tables \ref{Table1} and \ref{Table3}
that the critical temperature $T_{c}$ increases by decreasing the
power parameter for fixed $z$. As the power parameter $q$ becomes
larger, the critical temperature decreases. It means that the
magnetic moment is harder to be formed. This behavior has been
reported previously in Ref. \cite{Zhang2}, too. Figs. \ref{fig1}
and \ref{fig2} confirms the above results. Similarly, for a
constant value of $q$ and comparing the situation of $z=3/2$ with
$z=7/4$ for $d=4$ and $z=3/2$ with $z=17/10$ for $d=5$\footnote{
Since our chosen masses should satisfy the
Breitenlohner-Freedman(BF) bound, $4m^{2}-[z-(6-d)]^{2}>0$,
therefore by considering $m^{2}=1/8$, the dynamical exponent
should be confined with $(6-d)-2m<z<2m+(6-d)$. So we have chosen
the values for dynamical exponent as $z=3/2, 7/4$ for $d=4$. For
$d=5$, we have considered $z=3/2, 17/10$ for dynamical exponent.
Thus, for these choice of the parameter $m$ and $d$, the values of
Lifshitz dynamical exponent is restricted $z<2$.}, we see that the
critical temperature $T_{c}$ will decrease with the increasing
$z$, and the magnetic moment will become smaller for PM nonlinear
electrodynamics. This fact can be found from Figs.\ref{fig1} and
\ref{fig2}. It is found that the spontaneous condensate in the
absence of external magnetic field appears. The results are
consistent with those of mean field theory as can be seen in
Tables \ref{Table2} and \ref{Table4}. From these Figs.\ref{fig1}
and \ref{fig2}, we find that the increasing value of PM parameter
$q$ makes the magnetic moment smaller for fixed values of $z$,
which is similar to the cases of nonlinear electrodynamics. It is
easy to find that as $z$ increases, the magnetic moment decreases
for different values of $q$ in $d=4$, which is understood by
comparing Fig. \ref{fig1a} and \ref{fig1b}. One can find the same
results for $d=5$ from Figs. \ref{fig2a} and \ref{fig2b}. However
there is different behavior for magnetic moment in case of $q=3/4$
regardless of the values of $z$. It can be understood from these
figures that by increasing $z$, the magnetic moment increases.
Therefore, by increasing $z$, both the magnetic moment and the
critical temperature decrease for $q=1, 5/4$. However, for
$q=3/4$, we observe a different behavior in which by increasing
$z$, although the critical temperature decreases, the magnetic
moment increases as well.

The magnetic susceptibility density is one of the characteristic
properties of ferromagnetic material. Let us turn on the external
magnetic field to investigate the response of magnetic moment
$N$, which is described the static susceptibility density as
\begin{equation}
 \chi=\lim\limits_{B\to 0}\frac{\partial N}{\partial B}.
\end{equation}
This feature can be obtained based on the previous analysis which
has been discussed in Ref. \cite{Cai3} where it has been found that the
susceptibility obeys Curie-Weiss law,
\begin{equation}
\chi=\frac{C}{T+\theta},\text{ \ \ \ \ } T>T_{C},
\end{equation}
where $C$ and $\theta$ are two constants. For PM nonlinear
electrodynamics, Fig. \ref{fig3} shows the behavior of
susceptibility density near the critical temperature in $4D$  for
$q=3/4, 1$ and $z=3/2$ and $7/4$. The same calculation for $d=5$ is
 presented in Fig. \ref{fig4} for $q=3/4, 1, 5/4$ and $z=3/2$ and $17/10$.
 In the paramagnetic phase for all considered cases, we observe that
 when the temperature is reduced, the magnetic susceptibility increases for
the fixed nonlinearly parameter $q$ and also dynamical exponent
$z$.  Moreover, the magnetic susceptibility satisfies the
Curie-Weiss law of the ferromagnetism near the critical
temperature regardless of the value of $q$ in different dimensions.
 In the high temperature regime, one can see that when the temperature
decreases, $\chi$ increases, and the susceptibility obeys the
Curie-Weiss law. These results have been presented in Tables
\ref{table5} and \ref{table6}. Obviously we observe that, for
fixed allowed value of dynamical exponent $z$, the coefficient in
front of $T/T_{c}$ increases,when the power parameters $q$
increases too, while for fixed value of $q$, this coefficient
decreases with increasing $z$. Meanwhile, we see that the absolute
value of $\theta/\mu$ will decrease for $q=1,5/4$ by increasing
the dynamical exponent $z$. We find different behavior for
$q=3/4$, as it can be found from these tables, the absolute value
of $\theta/\mu$ increases by increasing the dynamical exponent $z$
(see Tables \ref{table5}and \ref{table6}).
\section{Conclusions}
To summarize, we studied the physical properties of holographic
paramagnetic-ferromagnetic phase transition in the background of
$d$-dimensional Lifshitz black holes. We considered the PM
electrodynamics, and obtained the effects of nonlinear power
parameter $q$ and dynamical Lifshitz exponent $z$ on the
phase behavior of the system. We performed numerical shooting
method for studying our holographic model. We observed that for
this kind of nonlinear electrodynamics, both the power parameter
and dynamical exponent can make the condensation harder to form,
and the critical temperature and magnetic moment decrease as well for
any dimension.
It was confirmed that the enhancement in power parameter of
electrodynamics and dynamical exponent, cause the paramagnetic
phase more difficult to be appeared. Our data confirms these
results. We observed that for fixed values of dynamical exponent,
the effect of increasing power parameter causes the lower values
for the critical temperature in our model. Besides, for smaller
values of the power parameter, the gap in the magnetic moment in
the absence of magnetic field is larger which in turn exhibits
that the condensation is formed easier. In a fixed value of power
parameter, by increasing the dynamical exponent, the
magnetic moment decreases which causes the condensation to be
formed harder. Although for $q=3/4$ in $d=5$ this behavior is
opposite. The behavior of the magnetic moment is always as
$(1-T/T_{c})^{1/2}$. This is in agreement with the result of
mean field theory. One can
conclude that the explicit form of nonlinear electrodynamics and
the anisotropy of spacetime with different dimensions do not have
any effect on the relationship. Moreover, in the presence of
external magnetic field, the inverse magnetic susceptibility near
the critical point behaves as ($\frac{C}{T+\theta}$) for different
values of power parameters with different allowed values of dynamical
exponent in different dimensions, and it satisfies the Curie- Weiss law.
The absolute value of $\theta$ decreases by increasing the dynamical
exponent for $q=1, 5/4$, while for $q=3/4$, the absolute value of $\theta/\mu$ increases by increasing the dynamical exponent $z$.
\begin{acknowledgments}
We are grateful to the referee for constructive comments which
helped us improve our paper. We thank the Research Council of
Shiraz University. The work of A.S. has been supported financially
by the Research Institute for Astronomy \& Astrophysics of Maragha
(RIAAM), Iran.
\end{acknowledgments}


\begin{thebibliography}{99}

\bibitem{Bardeen1} J. Bardeen, L.N. Cooper and J. R. Schrieffer, {\itshape %
    Microscopic Theory of Superconductivity}, Phys. Rev. \textbf{106}, 162
(1957).

\bibitem{Cooper2} J. Bardeen, L. N. Cooper and J. R. Schrieffer, {\itshape Theory
    of Superconductivity}, Phys. Rev. \textbf{108}, 1175 (1957).

\bibitem{Maldacena1} J. M. Maldacena, {\itshape The large-N limit of superconformal
field theories and supergravity}, Adv. Theor. Math. Phys. \textbf{2}, 231
(1998) [arXiv:9711200].

\bibitem{Gubser2} S. S. Gubser, I. R. Klebanov and A. M. Polyakov, {\itshape Gauge
theory correlators from non-critical string theory}, Phys. Lett. B \textbf{%
428}, 105 (1998) [arXiv:9802109].

\bibitem{Witten3} E. Witten, {\itshape Anti-de Sitter space and holography}, Adv.
Theor. Math. Phys. \textbf{2}, 253 (1998) [arXiv:9802150].

\bibitem{hartnoll} S. A. Hartnoll, {\itshape Lectures on holographic methods for
condensed matter physics}, Class. Quant. Grav. \textbf{26}, 224002 (2009)
[arXiv:0903.3246].

\bibitem{Herzog5} C. P. Herzog, {\itshape Lectures on Holographic Superfluidity
and Superconductivity}, J. Phys. A \textbf{42}, 343001 (2009)
[arXiv:0904.1975].

\bibitem{McGreevy6} J. McGreevy, {\itshape Holographic duality with a view toward
many-body physics}, Adv. High Energy Phys. \textbf{2010}, 723105 (2010)
[arXiv:0909.0518].

\bibitem{Herzog7} C. P. Herzog, {\itshape Analytic holographic superconductor},
Phys. Rev. D \textbf{81}, 126009 (2010) [arXiv:1003.3278].

\bibitem{Gubser8} S. S. Gubser, {\itshape Breaking an Abelian gauge symmetry near
a black hole horizon}, Phys. Rev. D \textbf{78}, 065034 (2008)
[arXiv:0801.2977]


\bibitem{montull} M. Montull, A. Pomarol and P. J. Silva,
 {\itshape The Holographic Superconductor Vortex}, Phys. Rev. Lett. \textbf{103}, 091601 (2009) [arXiv:0906.2396].

 \bibitem{Donos} A. Donos, J. P. Gauntlett, J. Sonner and B. Withers,
 {\itshape Competing orders in M-theory: superfluids, stripes and metamagnetism}, JHEP \textbf{1303}, 108 (2013) [arXiv:1212.0871].

 \bibitem{albash} T. Albash and C. V. Johnson,
 {\itshape A Holographic Superconductor in an External Magnetic Field}, JHEP \textbf{0809}, 121 (2008) [arXiv:0804.3466].

 \bibitem{m.pujo} M. Montull, O. Pujolas, A. Salvio and P. J. Silva,
 {\itshape Magnetic Response in the Holographic Insulator/Superconductor Transition}, JHEP \textbf{1204}, 135 (2012) [arXiv:1202.0006].

\bibitem{iqbal} N. Iqbal, H. Liu, M. Mezei and Q. Si,
{\itshape Quantum phase transitions in holographic models of magnetism and superconductors}, Phys. Rev. D \textbf{82}, 045002 (2010) [arXiv:1003.0010].

 \bibitem{Building Hartnoll} S.A. Hartnoll, C.P. Herzog, G.T. Horowitz,
 {\itshape Building a holographic superconductor}, Phys. Rev. Lett. \textbf{101}, 031601 (2008) [arXiv:0803.3295].

\bibitem{Lai} C. Lai, Q. Pan, J. Jing, Y. Wang,
  \textit{Analytical study on holographic superfluid in AdS soliton background}, Phys. Lett. B {\bf757}, 65 (2016), [arXiv:1601.00134].

\bibitem{Rogatko} M. Rogatko, K.I. Wysokinski,
\textit{Condensate flow in holographic models in the presence of
dark matter}, J. High Energy Phys. {\bf1610}, 152 (2016),
[arXiv:1608.00343].

\bibitem{Kuang} X.M. Kuang, E. Papantonopoulos,
\textit{Building a holographic superconductor with a scalar field coupled kinematically to Einstein tensor}
, J. High Energy Phys. {\bf1608}, 161 (2016), [arXiv:1607.04928]
[hep-th].

\bibitem{Mansoori} S.A.H. Mansoori, B. Mirza, A. Mokhtari, F.L. Dezaki, Z. Sherkatghanad,
\textit{Weyl holographic superconductor in the Lifshitz black hole
background}, JHEP {\bf1607}, 111 (2016), [arXiv:1602.07245].

\bibitem{Ling} Y. Ling, P. Liu, C. Niu, J.P. Wu, Z.Y. Xian,
\textit{Holographic entanglement entropy close to quantum phase
transitions,} JHEP {\bf1604}, 114 (2016), [arXiv:1502.03661].

\bibitem{Introduction Cai}R.G. Cai, L. Li, L.F. Li, R.Q. Yang,
\textit{ Introduction to holographic superconductor models}, Sci.
China, Phys. Mech. Astron. {\bf58} (6), 060401 (2015) ,
[arXiv:1502.00437].

\bibitem{Lifshitz Wu} Y.B. Wu, J.W. Lu, C.Y. Zhang, N. Zhang, X. Zhang, Z.Q. Yang, S.Y. Wu,
\textit{Lifshitz effects on holographic p-wave superfluid}, Phys.
Lett. B {\bf741}, 138 (2015), [arXiv:1412.3689].


\bibitem{Bahareh} B. Binaei Ghotbabadi, M. Kord Zangeneh, A. Sheykhi,
\textit{One-dimensional backreacting holographic superconductors with exponential nonlinear electrodynamics}, Eur. Phys. J. C \textbf{78}, 381 (2018), [arXiv:1804.05442].

\bibitem{Mahya1} M. Mohammadi, A. Sheykhi and M. Kord Zangeneh,
\textit{Analytical and numerical study of backreacting
one-dimensional holographic superconductors in the presence of
Born-Infeld electrodynamics}, Eur. Phys. J. C \textbf{78}, 654
(2018), [arXiv:1805.07377].

\bibitem{Mahya2} M. Mohammadi, A. Sheykhi and M. Kord Zangeneh, \textit{One-dimensional backreacting holographic p-wave superconductors}, Eur. Phys. J.
C \textbf{78}, 984 (2018), [arXiv:1901.10540].

\bibitem{Mahya3} M. Mohammadi, A. Sheykhi,
 \textit{Conductivity of the holographic p-wave superconductors with higher order corrections}, Eur. Phys. J.
C \textbf{79}, 473 (2019), [arXiv:1908.07992 ].

\bibitem{Mahya4}M. Mohammadi, A. Sheykhi,
textit{Conductivity of the one-dimensional holographic p-wave
superconductors in the presence of nonlinear electrodynamics},
Phys. Rev. D {\bf100}, 086012 (2019),[arXiv:1910.06082].


\bibitem{dyonic} R. G. Cai and R. Q. Yang,{\itshape Paramagnetism-Ferromagnetism Phase Transition in a Dyonic Black Hole}, Phys. Rev. D \textbf{90}, 081901 (2014) [arXiv:1404.2856].
\bibitem{p.Acai6} R. G. Cai and R. Q. Yang, {\itshape Holographic model for the paramagnetism/antiferromagnetism phase
transition}, Phys. Rev. D \textbf{91}, 086001 (2015) [arXiv:1404.7737].
\bibitem{Coexistence.Cai} R. G. Cai and R. Q. Yang, {\itshape Coexistence and competition of ferromagnetism and p-wave superconductivity
in holographic model}, Phys. Rev. D \textbf{91}, 026001(2015) [arXiv:1410.5080].
\bibitem{Yokoi} N. Yokoi, M. Ishihara, K. Sato and E. Saitoh,{\itshape Holographic realization of ferromagnets}, Phys. Rev. D \textbf{93}, 026002 (2016) [arXiv:1508.01626].
\bibitem{Cai3} R. G. Cai and R. Q. Yang, {\itshape Antisymmetric tensor field and spontaneous magnetization in holographic
duality}, Phys. Rev. D \textbf{92}, 046001 (2015) [arXiv:1504.00855].


\bibitem{Cai4} R. G. Cai, R. Q. Yang, Y. B. Wu and C. Y. Zhang ,{\itshape Massive 2-form field and holographic ferromagnetic
phase transition}, JHEP \textbf{021}, 1511 (2015) [arXiv:1507.00546].
\bibitem{Insulator.Cai} R. G. Cai and R. Q. Yang, {\itshape Insulator/metal phase transition and colossal magnetoresistance in holographic
model}, Phys. Rev. D \textbf{92}, 106002 (2015) [arXiv:1507.03105].
\bibitem{Understanding.Cai} R. G. Cai and R. Q. Yang,{\itshape Understanding strongly coupling magnetism from holographic duality} [arXiv:1601.02936].



\bibitem{Zhang2} C. Y. Zhang, Y. B. Wu, Y. N. Zhang, H. Y. Wang and M. M. Wu, {\itshape Holographic paramagnetism-ferromagnetism phase transition with the
nonlinear electrodynamics}, Nucl. Phys. B \textbf{914}, 446 (2017),
[arXiv:1609.09318v1].
\bibitem{Wu1} Y. B. Wu, C. Y. Zhang, J. W. Lu, B. Fan, S. Shu and Y. C. Liu, {\itshape Holographic paramagnetism-ferromagnetism
phase transition in the Born-Infeld electrodynamics}, Phys. Lett.
B \textbf{760}, 469 (2016).


\bibitem{PM1} J. Jing, Q. Pan, S. Chen, \textit{Holographic
superconductors with Power-Maxwell field}, JHEP {\bf11}, 045
(2011), [arXiv:1106.5181].

\bibitem{PM2} J. Jing, L. Jiang and Q. Pan,
\textit{Holographic superconductors for the Power-Maxwell field with backreactions}, Class. Quantum Grav. {\bf33}, 025001 (2016).


\bibitem{Shey1} A. Sheykhi, H. R. Salahi, A. Montakhab, \textit{Analytical and Numerical Study of Gauss-Bonnet Holographic
Superconductors with Power-Maxwell Field}, JHEP {\bf04}, 058
(2016) [arXiv:1603.00075].

\bibitem{Shey2} H. R. Salahi, A. Sheykhi, A. Montakhab, \textit{Effects of
Backreaction on Power-Maxwell Holographic Superconductors in
Gauss-Bonnet Gravity}, Eur. Phys. J. C \textbf{76} (2016) 575
[arXiv:1608.05025].

\bibitem{Shey3} A. Sheykhi,  F. Shamsi, S. Davatolhagh \textit{The upper critical magnetic field of holographic superconductor
with conformally invariant Power Maxwell electrodynamics}, Can. J.
Phys. \textbf{95}, 450 (2017), [arXiv:1609.05040].

\bibitem{Shey4} D. Hashemi Asl, A. Sheykhi,  \textit{Meissner-like effect and conductivity of power-Maxwell holographic
superconductors}, Phys. Rev. D {\bf101}, 026012 (2020),
[arXiv:1905.11810].

\bibitem{Doa} A. Sheykhi, D. Hashemi Asl, A. Dehyadegari,
\textit{Conductivity of higher dimensional holographic
superconductors with nonlinear electrodynamics}, Phys. Lett. B
\textbf{781}, 139 (2018) [arXiv:1803.05724].


\bibitem{Lif} S. Kachru, X. Liu, M. Mulligan,
\textit{Gravity Duals of Lifshitz-like Fixed Points}, Phys. Rev. D
\textbf{78}, 106005 (2008), [arXiv:0808.1725].


\bibitem {carlos} C. Hoyos, B. Soo Kim, Y. Oz, \textit{Lifshitz Hydrodynamics}, JHEP \textbf{2013}, 145 (2013), [arXiv:1304.7481].

\bibitem{Bu} Y. Bu, \textit{Holographic superconductors with z=2 Lifshitz scaling}, Phys. Rev. D \textbf{86}, 046007 (2012), [arXiv:1211.0037].




\bibitem{LU} J. W. Lu, Y. B Wu, P. Qian, Y. Y. Zhao, X. Zhang, N. Zhang, \textit{ Lifshitz Scaling Effects on Holographic Superconductors},
Nucl. Phys. B {\bf 887}, 112 (2014), [arXiv:1311.2699].

\bibitem{Natsuume} M. Natsuume and T. Okamura, \textit{Holographic Lifshitz superconductors: Analytic solution},
 Phys. Rev. D  \textbf{97}, 066016 (2018), [arXiv:1801.03154].


\bibitem{Sherkatghanad} Z. Sherkatghanad, B. Mirza, F. Lalehgani Dezaki,
\textit{Exponential nonlinear electrodynamics and backreaction effects on Holographic superconductor in the Lifshitz black hole background},
Int. J. Mod. Phys. D \textbf{26}, 1750175 (2017), [arXiv:1708.04289].


\bibitem{Zhao14} Z. Zhao, Q. Pan, J. Jing,
\textit{Notes on analytical study of holographic superconductors
with Lifshitz scaling in external magnetic field}, Phys. Lett. B
\textbf{735 }, 438 (2014), [arXiv:1311.6260].

\bibitem{Mahya5} M. Mohammadi, A. Sheykhi,
\textit{Lifshitz scaling effects on the holographic $p$-wave
superconductors coupled to nonlinear electrodynamics}, Eur. Phys.
J. C {\bf80}, 928 (2020), [arXiv:2010.07105].

\bibitem{Lifshitz5} C. Y. Zhang, Y. B. Wu, Y. Y. Jin, Y. T. Chai, M. H. Hu and Z. Zhang,
{\itshape Lifshitz scaling effects on the holographic
paramagnetism-ferromagnetism phase transition}, Phys. Rev. D
\textbf{93}, 126001 (2016) [arXiv:1603.04149].



\bibitem{Zhang2plb} Ya-Bo Wu, Cheng-Yuan Zhang, Jian-Bo Lu, Mu-Hong Hu,
 Yun-Tian Chai, {\itshape Holographic model for ferromagnetic phase transition
  in the Lifshitz black hole with the nonlinear electrodynamics},
  Phys. Lett. B \textbf{767}, 264 (2017).

\bibitem{binaei} B. Binaei Gh., A. Sheykhi, G. H. Bordbar,
 {\itshape Holographic paramagnetic ferromagnetic phase transition with Power Maxwell electrodynamics},
 Phys. Lett. B \textbf{ 797}, 134896 (2019) [arXiv:1903.05451].

\bibitem{shamsip30} M. Hassaine and C. Martinez, {\itshape Higher-dimensional black holes with a conformally invariant Maxwell source}, Phys. Rev. D
\textbf{75}, 027502 (2007) [arXiv:hep-th/0701058].

\bibitem{SheyPM} A. Sheykhi, \textit{Higher dimensional charged f(R) black holes}, Phys.\ Rev.\ D {\bf86}, 024013 (2012), [arXiv:1209.2960].

\bibitem{SheyPM2} A. Sheykhi, S.H. Hendi,
 \textit{Power-law entropic corrections to Newton law and
Friedmann equations} Phys.\ Rev.\ D {\bf84} 044023 (2011).

\bibitem{SheyPM3} A. Sheykhi and S. H. Hendi, \textit{Rotating black branes in
$f(R)$ gravity coupled to nonlinear Maxwell field}, Phys.\ Rev.\ D
{\bf87}, 084015 (2013).

\bibitem{dehyadegari} A. Dehyadegari, M. Kord Zangeneh and A. Sheykhi,{\itshape Holographic conductivity in the massive gravity with power-law Maxwell field}, Phys. Lett. B \textbf{773}, 344 (2017) [arXiv:1703.00975]


\bibitem{q.pan} Q. Pan, B. Wang, E. Papantonopoulos, J. Oliveira, A. B. Pavan,
{\itshape Holographic superconductors with various condensates in
Einstein-Gauss-Bonnet gravity}, Phys. Rev. D \textbf{81}, 106007
(2010) [arXiv:0912.2475].

\bibitem {Zhao} Z. Zhao, Q. Pan, S. Chen, J. Jing, {\itshape Notes on holographic superconductor models with the nonlinear electrodynamics},
Nucl. Phys. B \textbf{871}, 98 (2013) [arXiv:1212.6693]

\bibitem {Jing} J. Jing, Q. Pan, S. Chen,{\itshape Holographic superconductor/insulator transition with
 logarithmic electromagnetic field in Gauss-Bonnet gravity}, Phys. Lett. B \textbf{716}, 385 (2012) [arXiv:1209.0893].




\end{thebibliography}
\end{document}